\documentclass[aps,prd,amsmath,amssymb,superscriptaddress,preprintnumbers,twocolumn,nofootinbib,10pt]{revtex4-1}
\pdfoutput=1
\usepackage{graphicx}
\usepackage{dcolumn}
\usepackage{bm}
\usepackage{amssymb}
\usepackage{latexsym}
\usepackage{booktabs}
\usepackage{amsmath}
\usepackage{multirow}
\usepackage{url}
\usepackage{footnote}
\usepackage{float}
\usepackage{threeparttable}
\usepackage[colorlinks=true, linkcolor=blue, citecolor=blue]{hyperref}
\usepackage[bottom]{footmisc}

\usepackage[normalem]{ulem}
\usepackage{color}
\usepackage{array}
\usepackage{enumerate}
\usepackage{adjustbox}
\usepackage{makecell}
\usepackage{diagbox}
\usepackage{epstopdf}
\usepackage{epsfig}
\usepackage{longtable}
\usepackage{supertabular}
\usepackage{algorithm}
\usepackage{pifont}
\usepackage{algorithmic}
\usepackage{changepage}
\usepackage{setspace}
\usepackage{tabularx}
\newcolumntype{C}{>{\centering\arraybackslash}X}

\begin{document}

\title{Measuring neutrino masses with joint JWST and DESI DR2 data}

\author{Sheng-Han Zhou}
\affiliation{Liaoning Key Laboratory of Cosmology and Astrophysics, College of Sciences, Northeastern University, Shenyang 110819, China}

\author{Tian-Nuo Li}
\affiliation{Liaoning Key Laboratory of Cosmology and Astrophysics, College of Sciences, Northeastern University, Shenyang 110819, China}

\author{Guo-Hong Du}
\affiliation{Liaoning Key Laboratory of Cosmology and Astrophysics, College of Sciences, Northeastern University, Shenyang 110819, China}

\author{Jun-Qian Jiang}
\affiliation{School of Physical Sciences, University of Chinese Academy of Sciences, Beijing 100049, China}

\author{Jing-Fei Zhang}
\affiliation{Liaoning Key Laboratory of Cosmology and Astrophysics, College of Sciences, Northeastern University, Shenyang 110819, China}

\author{Xin Zhang}\thanks{Corresponding author}\email{zhangxin@mail.neu.edu.cn}
\affiliation{Liaoning Key Laboratory of Cosmology and Astrophysics, College of Sciences, Northeastern University, Shenyang 110819, China}
\affiliation{MOE Key Laboratory of Data Analytics and Optimization for Smart Industry, Northeastern University, Shenyang 110819, China}
\affiliation{National Frontiers Science Center for Industrial Intelligence and Systems Optimization, Northeastern University, Shenyang 110819, China}

\begin{abstract}
Early James Webb Space Telescope (JWST) observations reveal an unexpectedly abundant population of high-redshift candidate massive galaxies at $z \gtrsim 7$, and recent DESI measurements show a preference for dynamical dark energy, which together present a significant challenge to the standard $\Lambda$ cold dark matter ($\Lambda$CDM) cosmology. In this work, we jointly analyze high-redshift galaxy data from JWST, baryon acoustic oscillations data from DESI DR2, and cosmic microwave background (CMB) data from Planck and ACT, measuring the total neutrino mass $\sum m_{\nu}$. We consider three dark energy models ($\Lambda$CDM, $w$CDM, and $w_0w_a$CDM) and three mass hierarchies. Our results indicate that in the $w_0w_a$CDM model, adding JWST data to CMB+DESI tightens the upper limit of $\sum m_{\nu}$ by about $5.8\%-10.2\%$, and we obtain $\sum m_{\nu} < 0.167~\mathrm{eV}$ ($2\sigma$) in the normal hierarchy (NH) case. Furthermore, JWST also offers indicative lower limits on star formation efficiency parameter of $f_{*,10} \gtrsim 0.146-0.161$. Bayesian evidence weakly favors the $w_0w_a$CDM+$\sum m_{\nu}$(NH) model relative to the $\Lambda$CDM+$\sum m_{\nu}$(NH) model using CMB+DESI+JWST data. These results suggest that the joint analysis of high-redshift JWST data and low-redshift DESI data provides useful constraints on neutrino mass and merits further investigation.
\end{abstract}

\maketitle
\section{Introduction}\label{sec1}

The discovery of neutrino oscillations, establishing that neutrinos possess nonzero masses, has motivated extensive studies of neutrinos in the framework of physics beyond the Standard Model \cite{Super-Kamiokande:1998kpq,Super-Kamiokande:2000ywb,SNO:2001kpb,ParticleDataGroup:2014cgo,DayaBay:2022orm}. Neutrinos exist in three mass eigenstates ($\nu_1, \nu_2, \nu_3$) with masses ($m_1, m_2, m_3$) and three flavor eigenstates ($\nu_e, \nu_{\mu}, \nu_{\tau}$). Oscillation experiments measure the squared mass differences $\Delta m_{21}^2 \sim 7.54\times 10^{-5}\ {\mathrm{eV}^2}$ and $|\Delta m_{31}^2| \sim 2.45\times 10^{-3}\ {\mathrm{eV}^2}$ \cite{deSalas:2020pgw, Esteban:2024eli, Capozzi:2025wyn}, but cannot determine the absolute mass or the sign of $\Delta m_{31}^2$, leaving two possible hierarchies: normal hierarchy (NH, $m_1 < m_2 \ll m_3$) and inverted hierarchy (IH, $m_3 \ll m_1 < m_2$). When mass splittings are neglected, degenerate hierarchy (DH, $m_1 \approx m_2 \approx m_3$) is also considered \cite{Lesgourgues:2006nd}. Direct measurements for the absolute neutrino mass are pursued through two main approaches: $\beta$-decay experiments and neutrinoless double-$\beta$ decay experiments. The KATRIN $\beta$-decay experiment currently sets an upper limit on the total neutrino mass of $\sum m_{\nu} < 1.35~\mathrm{eV}$ \cite{KATRIN:2024cdt}. Meanwhile, the KamLAND-Zen neutrinoless double-$\beta$ decay experiment sets an upper limit of $\sum m_{\nu} < 0.12-0.39~\mathrm{eV}$ \cite{KamLAND-Zen:2024eml}.

Although oscillation experiments and direct measurements have made remarkable progress, cosmological observations are indispensable for measuring $\sum m_{\nu}$. This is because they probe the imprint of massive neutrinos on the cosmic microwave background (CMB) and large-scale structure, and currently provide the most stringent limits. To robustly measure $\sum m_{\nu}$ via cosmology requires baseline cosmological model. The standard $\Lambda$ cold dark matter ($\Lambda$CDM) model is composed of the cosmological constant representing dark energy and pressureless cold dark matter. Within this framework, cosmological probes already place stringent limits on $\sum m_{\nu}$. For instance, Planck CMB data alone yields $\sum m_{\nu} < 0.24~\mathrm{eV}$ ($2\sigma$) in the DH case \cite{Planck:2018vyg}, highlighting cosmological sensitivity to the absolute neutrino mass. 

However, $\Lambda$CDM model still faces theoretical problems such as the ``fine-tuning" and ``cosmic coincidence" problems \cite{Sahni:1999gb,Bean:2005ru}, as well as observational tensions: the $H_0$ tension\footnote{Local measurements of the Hubble constant $H_0$, primarily from the SH0ES collaboration using the Cepheid-calibrated distance ladder with type Ia supernova (SN), exhibit $5\sigma$ higher than the value inferred from Planck CMB observations within the $\Lambda$CDM model \cite{Planck:2018vyg,Riess:2021jrx}.} and the $S_8$ tension\footnote{The $S_8$ tension between weak-lensing surveys and Planck is alleviated in the updated Kilo-Degree Survey cosmic shear analysis, which reduces the discrepancy to $0.73\sigma$, rendering the result consistent with Planck \cite{Planck:2018vyg,Wright:2025xka}.}.
These two tensions have been extensively debated, with ongoing discussions about their statistical significance and possible physical origins, as discussed in Refs.~\cite{Guo:2018ans,Verde:2019ivm,DiValentino:2021izs,Vagnozzi:2023nrq}. These persistent issues motivate the exploration of physics beyond $\Lambda$CDM, such as holographic dark energy models \cite{Huang:2004wt,Zhang:2005hs}, interacting dark energy models \cite{Farrar:2003uw,Zhang:2004gc}, early dark energy models \cite{Poulin:2018cxd}, modifications of gravity \cite{Clifton:2011jh}, and dynamical dark energy models \cite{Boisseau:2000pr}. For a recent comprehensive review, see Ref.~\cite{CosmoVerseNetwork:2025alb}.

In recent years, measuring neutrino mass has been extensively explored within a variety of dark energy models \cite{Mainini:2010ng,Li:2012spm,Zhang:2014ifa,Zhang:2015uhk,Feng:2017usu,Vagnozzi:2018jhn,Guo:2018gyo,DiazRivero:2019ukx,Liu:2020vgn,Yao:2022jrw,Chudaykin:2022rnl,Sharma:2022ifr,Li:2023gtu,Jiang:2024viw,Barua:2025adv}. In particular, \citet{Zhang:2015uhk,Zhang:2017rbg} performed a comprehensive analysis across a wide range of dark energy models, showing that in dynamical dark energy scenarios the upper limit of $\sum m_{\nu}$ tends to shrink as dark-energy equation of state parameter $w$ shifts to smaller values. This conclusion has been further supported by some later studies \cite{Hannestad:2005gj,Wang:2016tsz,Yang:2017amu,RoyChoudhury:2018gay}.

Recently, several new observations have posed notable challenges to the $\Lambda$CDM model. The Dark Energy Spectroscopic Instrument (DESI) Data Release 2 (DR2), providing baryon acoustic oscillations (BAO) measurements from 14 million extragalactic sources, has delivered exceptionally precise low-redshift constraints on the expansion history and the growth of structure. When combined with CMB and SN data, DESI DR2 reveals a $2.8\sigma$-$4.2\sigma$ preference for the $w_0w_a$CDM model \cite{DESI:2025zgx}, and numerous studies have explored its broader implications for cosmology and fundamental physics \cite{Colgain:2024xqj,Li:2024qso,Jiang:2024xnu,Du:2024pai,Ye:2024ywg,Wu:2024faw,Li:2024qus,Li:2024bwr,Wang:2024dka,Giare:2024smz,Huang:2025som,Li:2025owk,Li:2025ula,Barua:2025ypw,Ling:2025lmw,Pang:2025lvh,Pan:2025qwy,Liu:2025myr,Yang:2025ume,Chen:2025wwn,Paliathanasis:2025xxm,Cai:2025mas,vanderWesthuizen:2025rip,Yang:2025oax,Yao:2025kuz,Pedrotti:2025ccw,Du:2025xes}. In addition, observations from the James Webb Space Telescope (JWST) have dramatically advanced the study of the early universe, revealing candidate massive galaxies already in place at $z \gtrsim 7$ \cite{Labbe:2022ahb}. The existence of such galaxies suggests star formation efficiency (SFE) significantly higher than anticipated, thereby challenging the $\Lambda$CDM model \cite{Menci:2022wia,Forconi:2023hsj,Alonso:2023oro,Wang:2023ros,Adil:2023ara,Lei:2023mke,Menci:2024rbq,Menci:2024hop,Jiang:2024tll,Shen:2024hpx,Wang:2024hce,Liu:2024yan,Chakraborty:2025yuo}.

Specifically, regarding neutrino mass measurements, the DESI Collaboration has reported an upper limit of $\sum m_{\nu} < 0.064$ eV (2$\sigma$) within the $\Lambda$CDM model with the combination of CMB and DESI DR2 data, considering the DH case \cite{DESI:2025zgx}. Additionally, since neutrinos affect cosmic structure formation and galaxy evolution through their free-streaming effects, JWST observations of massive high-redshift galaxies at $z \gtrsim 7$ provide a valuable method for measuring $\sum m_{\nu}$. In this context, combining JWST with Planck 2018 CMB data, \citet{Liu:2023qkf} reported an upper limit of $\sum m_{\nu} < 0.196$ eV ($2\sigma$), representing an 18\% improvement compared to CMB alone.

In this work, we measure $\sum m_{\nu}$ within $\Lambda$CDM, $w$CDM, and $w_0w_a$CDM models, considering three mass hierarchies: DH, NH, and IH. We perform observational constraints by jointly analyzing high-redshift galaxy data from JWST, the latest BAO measurements from DESI DR2, and CMB data from Planck and the Atacama Cosmology Telescope (ACT). In addition, we use Bayesian evidence to identify which dark energy models and neutrino mass hierarchies are supported by the latest observational data.

This paper is organized as follows. In Sec.{~\ref{sec2}}, we describe the theoretical analysis methods and datasets. In Sec.{~\ref{sec3}}, we report and discuss the main results. Finally, we present conclusions of this paper in Sec.{~\ref{sec4}}.

\section{Methodology and Data}\label{sec2}
\subsection{Models}
The mass splittings between the three active neutrino flavors are taken into account
\begin{equation}\label{1}
\Delta m_{21}^{2} \equiv m_{2}^{2} - m_{1}^{2} = 7.54\times 10^{-5}~\mathrm{eV}^{2},
\end{equation}
\begin{equation}\label{2}
|\Delta m_{31}^{2}| \equiv |m_{3}^{2} - m_{1}^{2}| = 2.46\times 10^{-3}~\mathrm{eV}^{2}.
\end{equation}

In the analysis of neutrino mass hierarchies, $\sum m_{\nu}$ is expressed as a function of the lightest mass eigenstate and the mass-squared differences from oscillation experiments. For the NH case, $\sum m_{\nu}$ is given by
\begin{equation}\label{3}
\sum m_{\nu}^{\mathrm{NH}} = m_1 + \sqrt{m_1^2 + \Delta m_{21}^2} + \sqrt{m_1^2 +|\Delta m_{31}^2|},
\end{equation}
where $m_1$ denotes the lightest neutrino mass. Here, $\Delta m_{21}^2 \equiv m_2^2 - m_1^2$ and $\Delta m_{31}^2 \equiv m_3^2 - m_1^2$. For the IH case, the expression is
\begin{equation}\label{4}
\sum m_{\nu}^{\mathrm{IH}} = m_3 + \sqrt{m_3^2 + |\Delta m_{31}^2|} + \sqrt{m_3^2 + |\Delta m_{31}^2| + \Delta m_{21}^2},
\end{equation}
with $m_3$ being the lightest mass and $|\Delta m_{31}^2|$ accounting for the negative $\Delta m_{31}^2$ in IH. For the DH case, mass splittings are neglected, yielding
\begin{equation}\label{5}
\sum m_{\nu}^{\mathrm{DH}} = 3m,
\end{equation}
where $m$ is the free mass parameter. Lower limits from oscillation data are imposed as priors: $\sum m_{\nu} > 0\ ~\mathrm{eV}$ (DH), $\sum m_{\nu} > 0.06\ ~\mathrm{eV}$ (NH), and $\sum m_{\nu} > 0.10\ ~\mathrm{eV}$ (IH).

In cosmology, the gravitational effect of the three standard neutrino species is incorporated by including $\sum m_{\nu}$ as a component of the total matter density. The fractional contribution of neutrinos to the critical density of the universe is given by \cite{Lesgourgues:2006nd}
\begin{equation}
\Omega_{\nu} = \frac{\rho_\nu}{\rho_{\rm crit}} = \frac{\sum m_{\nu}}{93.14\,h^2~{\mathrm{eV}}}~,
\end{equation}
where $\rho_\nu$ is the present total energy density of neutrinos, $\rho_{\rm crit} = 3H_0^2/(8\pi G)$ is the critical density of the universe, and $h \equiv H_0 / (100~\mathrm{km~s^{-1}~Mpc^{-1}})$ is the dimensionless Hubble constant. This expression follows from the fact that neutrinos behave as relativistic species in the early universe and transition to nonrelativistic matter by the present day, with their energy density taking the form given above.

In this work, we adopt the $\Lambda\mathrm{CDM}+\sum m_{\nu}$ model as the baseline cosmology, extend it to the $w\mathrm{CDM}+\sum m_{\nu}$ scenario for comparison, and further explore the $w_0w_a\mathrm{CDM}+\sum m_{\nu}$ framework as our primary focus.

\subsection{Likelihood of JWST}
The excess number of massive, high-redshift galaxies can place constraints on the SFE, thereby limiting viable cosmological models. \citet{Wang:2023gla} allows for a rapid estimation of the tension between cosmological models and the high-redshift massive galaxy candidates published by the JWST. \citet{Liu:2023qkf} extended the likelihood framework to accommodate massive neutrinos\footnote{\url{http://zhiqihuang.top/codes/massivehalo.htm}.}.
The halo mass function is assumed to take the following form,
\begin{equation}
\frac{\mathrm{d}N}{\mathrm{d}M} = \frac{\bar{\rho}}{M^2}\frac{\mathrm{d}\ln \nu}{\mathrm{d}\ln M}f(\nu). \label{eq:haloMF}
\end{equation}

Here, $\bar{\rho}$ denotes the average background density, with $\nu = \delta_\mathrm{c} / \sigma(M,z)$. In this context, $\delta_\mathrm{c} = 1.686$ represents the critical linear overdensity, and $\sigma(M,z)$ gives the mass fluctuation at scale $R = (3M / 4\pi \bar{\rho})^{1/3}$. 

The simulation-calibrated $f(\nu)$ factor is given by
\begin{equation}
\nu f(\nu) = A (\frac{1}{(\sqrt{a}\nu)^{q}}+1) (\frac{\sqrt{a}\nu}{2})^{\frac{{1}}{2}} \frac{e^{\frac{-a \nu^{2}}{2}}}{\sqrt{\pi}},
\end{equation}
where $a=0.707$, $A=0.322$ and $q=0.3$ \cite{Sheth:1999mn}. For massive dark matter halos, $f_\mathrm{b} \equiv \Omega_{\mathrm{b}} / \Omega_{\mathrm{m}}$ defines the baryonic mass fraction; here $\Omega_{\mathrm{b}}$ and $\Omega_{\mathrm{m}}$ are the baryon and total matter density parameters, respectively. The stellar mass $M_*$ is linked to the halo mass $M_{\rm halo}$ by $M_{\rm *} = \epsilon f_\mathrm{b} M_{\rm halo}$, with $\epsilon$ signifying the SFE. We adopt a pivot-mass parametrization
\begin{equation}
\epsilon(M_{\rm halo}) = f_{\star,10}\left(\frac{M_{\rm halo}}{10^{10} M_\odot}\right)^{\alpha_{\star,10}}.
\end{equation}
Here $f_{\star,10}$ is the SFE evaluated at the pivot mass ($10^{10}\,M_\odot$). The parameter $\alpha_{\star,10}$ controls the local mass dependence of the SFE around the pivot. In our baseline we set $\alpha_{\star,10}=0$ because the single-threshold counts over a narrow redshift range provide little leverage on this slope and it is degenerate with $f_{\star,10}$ and cosmology; fixing it yields a conservative minimal model. We treat $f_{\star,10}$ as a free amplitude parameter and sample it jointly with the cosmological parameters.

Consequently, a threshold in stellar mass, $M_{\rm *,cut}$, can be converted to an equivalent halo mass threshold using 
\begin{equation}
M_{\rm halo, cut} = \frac{M_{\rm *,cut}}{\epsilon f_\mathrm{b}}. 
\end{equation}

If we assume every halo above this threshold contains one massive central galaxy, then the number of galaxies with stellar mass above $M_{\rm *,cut}$ expected within a specific comoving volume is
\begin{equation}
\langle N_{\rm th} \rangle = 4\pi f_{\rm sky}\int_{M_{\rm halo, cut}}^{\infty}\mathrm{d}M \int_{z_{\min}}^{z_{\max}} \frac{\mathrm{d}n}{\mathrm{d}M} \frac{\mathrm{d}V}{\mathrm{d}z\mathrm{d}\Omega} \,\mathrm{d}z, \label{eq:Nth}
\end{equation}
where the selected comoving volume is defined by the redshift interval $[z_{\min}, z_{\max}]$ and the sky fraction $f_{\rm sky}$. Following Ref.~\cite{Wang:2023gla}, we take $z_{\min}=7$ and $z_{\max} = 10$. The survey area is $38\,\mathrm{arcmin}^2$, which gives $f_{\rm sky}=2.56\times 10^{-7}$. 
The comoving volume per redshift interval per solid angle, $\frac{\mathrm{d}V}{\mathrm{d}z\mathrm{d}\Omega}$, is specified by the cosmology. For spatially flat $\Lambda$CDM model, we have 
\begin{equation}
\frac{\mathrm{d}V}{\mathrm{d}z\mathrm{d}\Omega} = \frac{c^3}{H(z)} \left( \int_0^z \frac{\mathrm{d}z'}{H(z')} \right)^2,
\end{equation}
where $c$ is the speed of light and $H(z)$ is the Hubble parameter. By marginalizing over the cosmic variance of $\lambda$ (characterized by $\sigma_\lambda$, estimated following Trenti and Stiavelli~\cite{Trenti:2007dh}, using the public cosmic-variance emulator with a $\sim1.4$ nonlinear correction appropriate for pencil-beam survey volumes), we derive the distribution function of $N_{\mathrm{th}}$,
\begin{equation}
{P}\left(N_{\rm th}\right)=\int_{-\infty}^{\infty} \frac{1}{\sqrt{2\pi}\sigma_{\lambda}} e^{-\frac{(\lambda-\left \langle N_{\rm th} \right \rangle)^2}{2\sigma^2_\lambda}} e^{-\lambda} \frac{\lambda^{N_{\rm th}}}{N_{\rm th}!} \mathrm{d}\lambda.
\end{equation}

The likelihood of the theory is then characterized by the probability of finding $ N_{\mathrm{obs}} \leq N_{\mathrm{th}} $, which is given by
\begin{equation}
P(N_{\mathrm{obs}} \leq N_{\mathrm{th}}) = \sum_{N_{\mathrm{obs}}=0}^{\infty} P(N_{\mathrm{obs}}) \sum_{N_{\mathrm{th}}=N_{\mathrm{obs}}}^{\infty} P(N_{\mathrm{th}}).
\end{equation}

The distribution function $P(N_{\mathrm{th}})$ is set by the cosmological model and the SFE (with the Poisson mean marginalized over cosmic variance as above). For the data side, we compute $P(N_{\mathrm{obs}})$ from per-candidate inclusion probabilities: for each candidate we evaluate the probability of passing the stellar-mass and redshift cuts by marginalizing its reported $(\log M_\ast, z)$ posteriors (including simple systematics). Assuming independence across candidates, $N_{\mathrm{obs}}$ follows a Poisson-binomial distribution with parameters $\{p_i\}$; we obtain its probability mass function via a short dynamic-programming recursion. See Ref.~\cite{Wang:2023gla} for a detailed description.

\subsection{Cosmological datasets}
In this work, we utilize the following three cosmological observational datasets:

\begin{itemize}

\item \textit{Cosmic microwave background}. 
This analysis incorporates measurements from Planck CMB temperature anisotropy, polarization power spectra, their cross-spectra, and the joint ACT-Planck lensing power spectrum. The CMB likelihoods comprise four distinct components:

(i) Small-scale ($\ell > 30$) temperature and polarization anisotropy spectra ($C_{\ell}^{TT}$, $C_{\ell}^{TE}$, $C_{\ell}^{EE}$) derive from the Planck PR4 \texttt{CamSpec} likelihood \cite{Planck:2018vyg,Efstathiou:2019mdh,Rosenberg:2022sdy}.

(ii) Large-scale ($2 \leq \ell \leq 30$) temperature anisotropy power spectrum ($C_{\ell}^{TT}$) utilizes the Planck \texttt{Commander} likelihood \cite{Planck:2018vyg,Planck:2019nip}.

(iii) Large-scale ($2 \leq \ell \leq 30$) E-mode polarization spectrum ($C_{\ell}^{EE}$) employs the Planck \texttt{SimAll} likelihood \cite{Planck:2018vyg,Planck:2019nip}.

(iv) The CMB lensing likelihood, with the latest and most precise data coming from the combination of the NPIPE PR4 Planck CMB lensing reconstruction\footnote{\url{https://github.com/carronj/planck_PR4_lensing}.} \cite{Carron:2022eyg} and Data Release 6 of the ACT\footnote{\url{https://github.com/ACTCollaboration/act_dr6_lenslike}.} \cite{ACT:2023dou}. This combined likelihood suite is designated ``CMB."

\item \textit{Baryon acoustic oscillations}. 
The BAO measurements from DESI DR2 include tracers of the bright galaxy sample, luminous red galaxies, emission line galaxies, quasars, and the Lyman-$\alpha$ forest. These tracers are described through the transverse comoving distance $D_{\mathrm{M}}/r_{\mathrm{d}}$, the angle-averaged distance $D_{\mathrm{V}}/r_{\mathrm{d}}$, and the Hubble horizon $D_{\mathrm{H}}/r_{\mathrm{d}}$, where $r_{\mathrm{d}}$ is the comoving sound horizon at the drag epoch. We use DESI DR2 BAO measurements summarized in Table IV on Ref.~\cite{DESI:2025zgx}\footnote{\url{https://data.desi.lbl.gov/doc/releases}.}. We denote this full dataset as ``DESI."

\item \textit{High-Redshift massive galaxy candidates}. 
New data from the JWST have revealed a number of high-redshift galaxy candidates with unexpectedly high stellar masses. This means that at high redshifts ($z \gtrsim 7$), the SFE is higher than predicted by the $\Lambda\mathrm{CDM}$ model. To evaluate the tension between this observation and predictions within a given cosmological framework, we employ the updated sample of massive galaxy candidates presented in Ref.~\cite{Liu:2023qkf}, which builds upon the initial identification presented in Ref.~\cite{Labbe:2022ahb}. This suite of infrared data, hereafter referred to as ``JWST," is summarized in Table~\ref{tab1}.
\end{itemize}

\begin{table}[!htb]
\setlength\tabcolsep{14pt}
\renewcommand{\arraystretch}{1.5}
\centering
\caption{\label{tab1}Galaxy sample (11 objects) with confirmed redshifts ($z\in (7,10)$) and derived stellar masses.} 
\resizebox{0.4\textwidth}{!}{%
\begin{tabular}{ccc}
\hline
\hline
id & $z$  & lg($M_*/M_\odot$) \\
\hline
2859  & $8.1056^{+0.4925}_{-1.4928}$ & $10.0291^{+0.2375}_{-0.2656}$  \\
7274  & $7.7740^{+0.0530}_{-0.0574}$ & $9.8661^{+0.0942}_{-0.0574}$  \\
11184 & $7.3180^{+0.2837}_{-0.3451}$ & $10.1810^{+0.1020}_{-0.1035}$  \\
14924 & $8.8308^{+0.1685}_{-0.0899}$ & $10.0150^{+0.1557}_{-0.1376}$  \\
16624 & $8.5168^{+0.1940}_{-0.2174}$ & $9.2986^{+0.2732}_{-0.2393}$  \\
21834 & $8.5432^{+0.3175}_{-0.5131}$ & $9.6077^{+0.2632}_{-0.3241}$  \\
25666 & $7.9310^{+0.0950}_{-0.1637}$ & $9.5218^{+0.2289}_{-0.1000}$  \\
28984 & $7.5415^{+0.0820}_{-0.1405}$ & $9.5723^{+0.1261}_{-0.1496}$  \\
35300 & $7.7690^{+0.0030}_{-0.0030}$ & $10.3968^{+0.1852}_{-0.2266}$  \\
38094 & $7.4773^{+0.0440}_{-0.0417}$ & $10.8869^{+0.0948}_{-0.0817}$  \\
39575 & $7.9932^{+0.0006}_{-0.0006}$ & $9.3289^{+0.4309}_{-0.3949}$  \\
\hline
\end{tabular}%
}
\end{table}

\begin{table}[t]
\caption{Flat priors on the main cosmological parameters constrained in this paper.}
\begin{center}
\renewcommand{\arraystretch}{1.1}
\begin{tabular}{@{\hspace{0.5cm}} c @{\hspace{1.0cm}} c @{\hspace{0.8cm}}}
\hline
\hline
Parameter       & Prior \\
\hline
$\Omega_{\rm b} h^2$              & $\mathcal{U}[0.005,\,0.1]$ \\
$\Omega_{\rm c} h^2$              & $\mathcal{U}[0.01,\,0.99]$ \\
$\tau$                            & $\mathcal{U}[0.01,\,0.8]$ \\
$\ln 10^{10}A_{\rm s}$          & $\mathcal{U}[1.61,\,3.91]$ \\
$\theta_{\rm s}$                  & $\mathcal{U}[0.5,\,10]$ \\
$n_{\rm s}$                       & $\mathcal{U}[0.8,\,1.2]$ \\
$w~{\rm or}~w_0$                  & $\mathcal{U}[-3,\,1]$ \\
$w_a$                             & $\mathcal{U}[-3,\,2]$ \\
$\sum m_{\nu}$ (DH)           & $\mathcal{U}[0,\,5]$ \\
$\sum m_{\nu}$ (NH)           & $\mathcal{U}[0.06,\,5]$ \\
$\sum m_{\nu}$ (IH)           & $\mathcal{U}[0.1,\,5]$ \\
$f_{\ast,10}$                    & $\mathcal{U}[0.05,\,0.3]$ \\
\hline
\end{tabular}
\label{tab2}
\end{center}	
\end{table}

\begin{table*}[t]
\renewcommand\arraystretch{1.7}
\centering
\caption{The constraints on cosmological parameters from the current cosmological datasets. For $\sum m_{\nu}$ and the SFE parameter $f_{*,10}$ (when JWST data are included), we report the $2\sigma$ confidence level. The results for the other parameters are given as the $1\sigma$ confidence level. Here, $H_0$ is expressed in units of $\mathrm{km\,s^{-1}\,Mpc^{-1}}$.}
\label{tab3}
\resizebox{\textwidth}{!}{%
\large
\begin{tabular}{cccccccccc}
\toprule[1pt]
\toprule[1pt]
& \multicolumn{3}{c}{$~~\Lambda\mathrm{CDM}+\sum m_{\nu}~~$} & \multicolumn{3}{c}{$~w\mathrm{CDM}+\sum m_{\nu}~$} & \multicolumn{3}{c}{$w_0w_a\mathrm{CDM}+\sum m_{\nu}$} \\
\cmidrule[0.5pt](l{2pt}r{2pt}){2-4} \cmidrule[0.5pt](l{2pt}r{2pt}){5-7} \cmidrule[0.5pt](l{2pt}r{2pt}){8-10}
& DH & NH & IH & DH & NH & IH & DH & NH & IH \\
\midrule[1pt]
\multicolumn{10}{l}{\textbf{CMB}} \\
$H_0$ & $66.90^{+1.10}_{-0.58}$ & $66.45^{+0.94}_{-0.56}$ & $66.06^{+0.86}_{-0.54}$ & $> 82.60$ & $> 83.70$ & $> 84.30$ & $> 81.30$ & $> 82.90$ & $> 82.70$ \\
$S_8$ & $0.8322\pm 0.0096$ & $0.8330\pm 0.0100$ & $0.8334\pm 0.0098$ & $0.7680^{+0.0200}_{-0.0380}$ & $0.7650^{+0.0180}_{-0.0350}$ & $0.7620^{+0.0170}_{-0.0340}$ & $0.7710^{+0.0210}_{-0.0400}$ & $0.7670^{+0.0180}_{-0.0370}$ & $0.7660^{+0.0180}_{-0.0380}$ \\
$w/w_0$ & -- & -- & -- & $-1.61^{+0.18}_{-0.35}$ & $-1.66^{+0.18}_{-0.33}$ & $-1.70^{+0.17}_{-0.33}$ & $-1.36^{+0.42}_{-0.53}$ & $-1.43^{+0.40}_{-0.50}$ & $-1.45^{+0.41}_{-0.51}$ \\
$w_a$ & -- & -- & -- & -- & -- & -- & $< -0.409$ & $< -0.474$ & $< -0.417$ \\
$f_{*,10}$ & -- & -- & -- & -- & -- & -- & -- & -- & -- \\
$\sum m_{\nu}$ [$\mathrm{eV}$] & $< 0.216$ & $< 0.243$ & $< 0.264$ & $< 0.241$ & $< 0.286$ & $< 0.310$ & $< 0.247$ & $< 0.296$ & $< 0.309$ \\
\midrule[0.8pt]
\multicolumn{10}{l}{\textbf{CMB+JWST}} \\
$H_0$ & $67.00^{+1.00}_{-0.56}$ & $66.50^{+0.80}_{-0.61}$ & $66.15^{+0.74}_{-0.57}$ & $> 83.20$ & $> 83.70$ & $> 82.70$ & $> 81.80$ & $> 83.80$ & $> 84.20$ \\
$S_8$ & $0.8330\pm 0.0100$ & $0.8344\pm 0.0098$ & $0.8340\pm 0.0100$ & $0.7660^{+0.0200}_{-0.0340}$ & $0.7640^{+0.0190}_{-0.0350}$ & $0.7660^{+0.0210}_{-0.0340}$ & $0.7710^{+0.0200}_{-0.0400}$ & $0.7660^{+0.0180}_{-0.0370}$ & $0.7640^{+0.0180}_{-0.0340}$ \\
$w/w_0$ & -- & -- & -- & $-1.61^{+0.19}_{-0.31}$ & $-1.66^{+0.18}_{-0.32}$ & $-1.66^{+0.20}_{-0.33}$ & $-1.38^{+0.37}_{-0.53}$ & $-1.44^{+0.40}_{-0.49}$ & $-1.46\pm 0.40$ \\
$w_a$ & -- & -- & -- & -- & -- & -- & $< -0.333$ & $< -0.389$ & $< -0.543$ \\
$f_{*,10}$ & $> 0.154$ & $> 0.153$ & $> 0.161$ & $> 0.157$ & $> 0.153$ & $> 0.147$ & $> 0.156$ & $> 0.159$ & $> 0.158$ \\
$\sum m_{\nu}$ [$\mathrm{eV}$] & $< 0.196$ & $< 0.218$ & $< 0.240$ & $< 0.209$ & $< 0.255$ & $< 0.294$ & $< 0.221$ & $< 0.273$ & $< 0.277$ \\
\midrule[1pt]
\multicolumn{10}{l}{\textbf{CMB+DESI}} \\
$H_0$ & $68.37\pm 0.29$ & $68.12\pm 0.28$ & $67.93\pm 0.28$ & $69.30\pm 0.93$ & $69.67\pm 0.91$ & $69.86\pm 0.93$ & $63.70^{+1.90}_{-2.10}$ & $63.40^{+1.50}_{-2.20}$ & $63.20^{+1.40}_{-2.00}$ \\
$S_8$ & $0.8184\pm 0.0073$ & $0.8127\pm 0.0073$ & $0.8088\pm 0.0073$ & $0.8187\pm 0.0076$ & $0.8130\pm 0.0074$ & $0.8092\pm 0.0073$ & $0.8430\pm 0.0110$ & $0.8430\pm 0.0110$ & $0.8410^{+0.0110}_{-0.0100}$ \\
$w/w_0$ & -- & -- & -- & $-1.040\pm 0.037$ & $-1.066\pm 0.037$ & $-1.081^{+0.039}_{-0.035}$ & $-0.430\pm 0.220$ & $-0.380^{+0.250}_{-0.170}$ & $-0.350^{+0.230}_{-0.150}$ \\
$w_a$ & -- & -- & -- & -- & -- & -- & $-1.73\pm 0.64$ & $-1.92^{+0.50}_{-0.80}$ & $-2.04^{+0.43}_{-0.74}$ \\
$f_{*,10}$ & -- & -- & -- & -- & -- & -- & -- & -- & -- \\
$\sum m_{\nu}$ [$\mathrm{eV}$] & $< 0.057$ & $< 0.106$ & $< 0.137$ & $< 0.076$ & $< 0.125$ & $< 0.156$ & $< 0.154$ & $< 0.186$ & $< 0.195$ \\
\midrule[0.8pt]
\multicolumn{10}{l}{\textbf{CMB+DESI+JWST}} \\
$H_0$ & $68.37\pm 0.28$ & $68.12\pm 0.29$ & $67.93\pm 0.29$ & $69.34\pm 0.93$ & $69.66\pm 0.94$ & $69.84\pm 0.97$ & $63.90^{+1.70}_{-2.20}$ & $63.40^{+1.50}_{-2.20}$ & $63.10^{+1.50}_{-2.00}$ \\
$S_8$ & $0.8188\pm 0.0071$ & $0.8134\pm 0.0072$ & $0.8094\pm 0.0072$ & $0.8191\pm 0.0075$ & $0.8138\pm 0.0074$ & $0.8099\pm 0.0074$ & $0.8430^{+0.0120}_{-0.0110}$ & $0.8430\pm 0.0110$ & $0.8420\pm 0.0110$ \\
$w/w_0$ & -- & -- & -- & $-1.042\pm 0.037$ & $-1.065^{+0.039}_{-0.034}$ & $-1.080\pm 0.039$ & $-0.450\pm 0.210$ & $-0.390^{+0.240}_{-0.180}$ & $-0.340^{+0.230}_{-0.160}$ \\
$w_a$ & -- & -- & -- & -- & -- & -- & $-1.65\pm 0.63$ & $-1.87^{+0.56}_{-0.69}$ & $-2.07^{+0.39}_{-0.76}$ \\
$f_{*,10}$ & $> 0.149$ & $> 0.150$ & $> 0.154$ & $> 0.151$ & $> 0.153$ & $> 0.146$ & $> 0.153$ & $> 0.154$ & $> 0.149$ \\
$\sum m_{\nu}$ [$\mathrm{eV}$] & $< 0.055$ & $< 0.105$ & $< 0.137$ & $< 0.075$ & $< 0.121$ & $< 0.152$ & $< 0.145$ & $< 0.167$ & $< 0.184$ \\
\bottomrule[1pt]
\end{tabular}
}
\end{table*}

\begin{figure*}[htbp]
\begin{center}
\includegraphics[width=16cm]{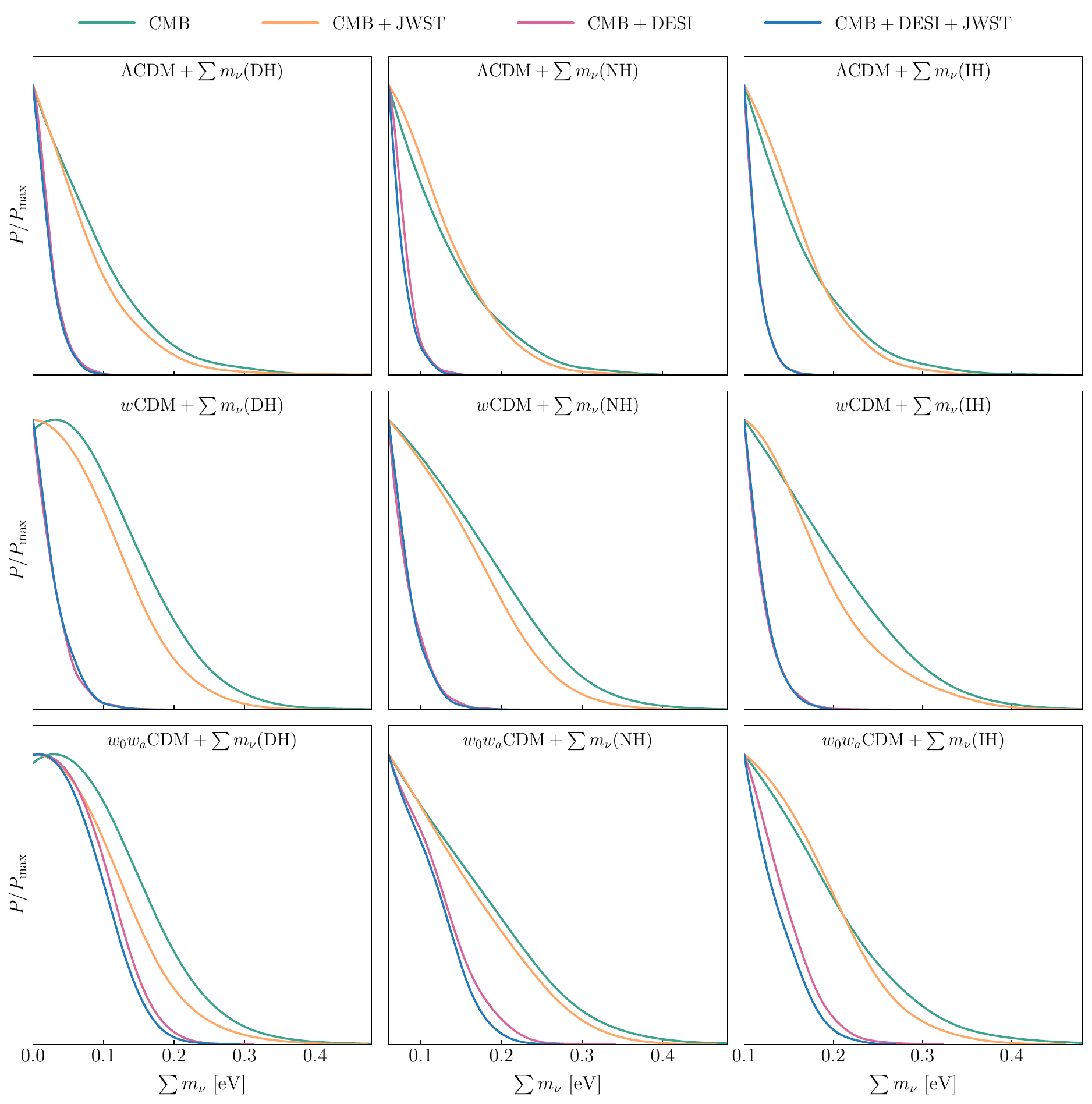}
\end{center}
\caption{The one-dimensional marginalized posterior distributions of $\sum m_{\nu}$ for the $\Lambda\mathrm{CDM}+\sum m_{\nu}$, $w\mathrm{CDM}+\sum m_{\nu}$, and $w_{0}w_{a}\mathrm{CDM}+\sum m_{\nu}$ models, based on the CMB, DESI, and JWST data.}
\label{figure1}
\end{figure*}

\begin{figure*}[htbp]
\begin{center}
\includegraphics[width=16.5cm]{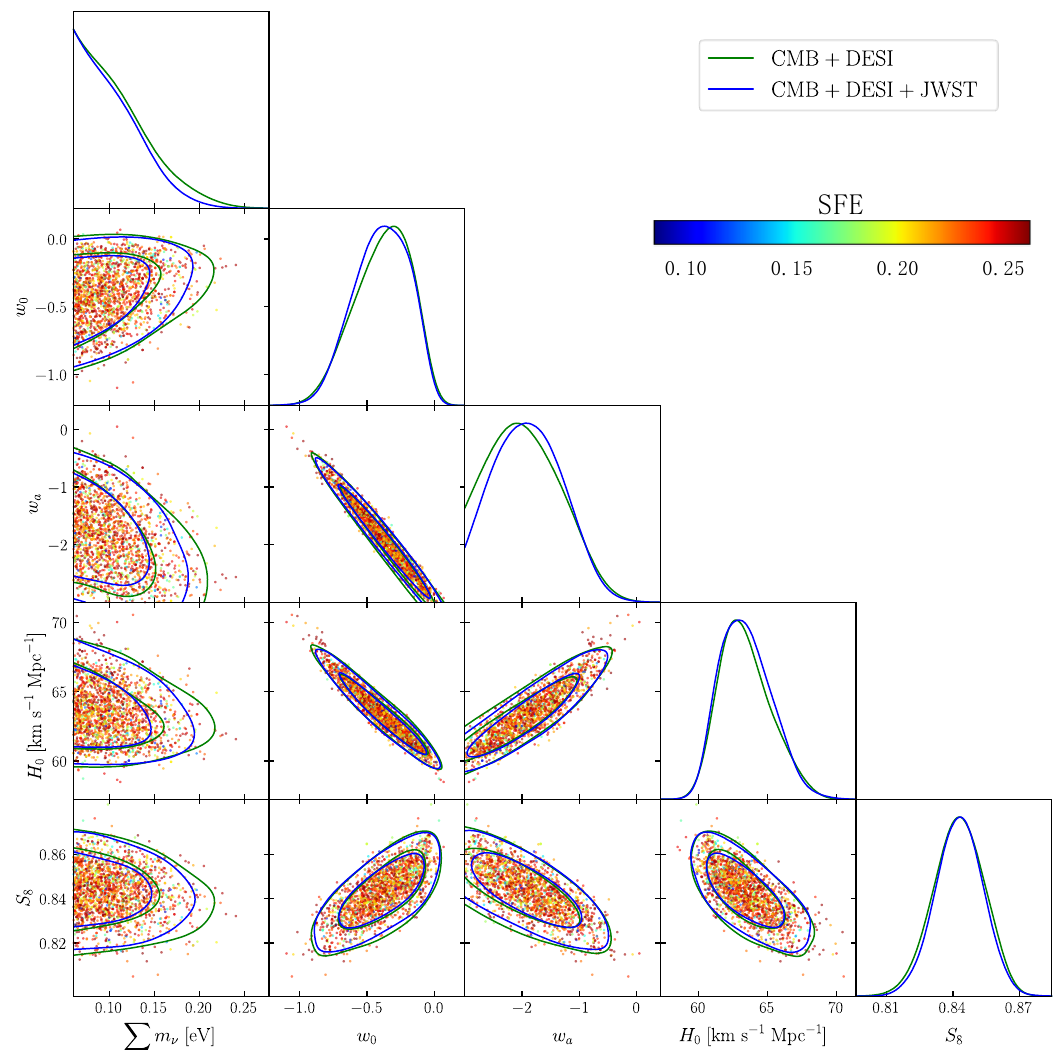}
\end{center}
\caption{Constraints on cosmological parameters in the $w_{0}w_{a}\mathrm{CDM}+\sum m_{\nu}$ (NH) model, using the CMB+DESI (green) and CMB+DESI+JWST (blue) datasets. The overlaid colored scatter points indicate the SFE parameter values, which are shown only for the CMB+DESI+JWST data.}
\label{figure2}
\end{figure*}

\subsection{Bayesian analysis}
In this paper, the theoretical model is implemented using the Boltzmann solver {\tt CAMB}\footnote{\url{https://github.com/cmbant/CAMB}.} \cite{Lewis:1999bs,Howlett:2012mh}. Bayesian inference employs the cosmological inference code {\tt Cobaya}\footnote{\url{https://github.com/CobayaSampler/cobaya}.} \cite{Torrado:2020dgo}, which is used to perform Markov chain Monte Carlo (MCMC) analyses. Table~\ref{tab2} summarizes the free parameters of the models along with the adopted uniform priors. The SFE parameter, $f_{*,10}$, quantifies the fraction of baryons converted into stars within a $10^{10} M_\odot$ halo. It is included only in analyses incorporating JWST data, with a flat prior of $0.05 \leq f_{*,10} \leq 0.3$ \cite{Bregman:2007ac,Wang:2025mdw}. The convergence of the MCMC chains is assessed using the Gelman-Rubin statistic with a criterion of $R - 1 < 0.02$, and the results are analyzed with the public package {\tt GetDist}\footnote{\url{https://github.com/cmbant/getdist}.} \cite{Lewis:2019xzd}.

The evaluation of different neutrino hierarchies and dark energy models in the light of the data is based on Bayesian evidence
\begin{equation}
p(d|\mathcal{M}) \equiv \int_{\Omega} p(d|\theta, \mathcal{M}) \, p(\theta |\mathcal{M}) \, \mathrm{d}\theta,
\end{equation}
where $ p(d|\mathcal{M}) $ is the marginal likelihood (also known as Bayesian evidence) for model $\mathcal{M}$, representing the probability of observing data $d$ under the model $\mathcal{M}$ after integrating over all possible parameter values $\theta$. Here, $\theta$ denotes the set of parameters characterizing the model $\mathcal{M}$, and $\Omega$ is the corresponding parameter space. 

Therefore, the posterior probability of a model $\mathcal{M}$ given the data $d$ can be obtained, which inverts the conditioning between data and model
\begin{equation}
p(\mathcal{M}|d) \propto p(\mathcal{M}) \, p(d|\mathcal{M}),
\end{equation}

When comparing two models, $\mathcal{M}_i$ versus $\mathcal{M}_j$, one is interested in the ratio of the posterior probabilities, or posterior odds, given by
\begin{equation}
\frac{p(\mathcal{M}_i|d)}{p(\mathcal{M}_j|d)} = \mathcal{B}_{ij}\frac{p(\mathcal{M}_i)}{p(\mathcal{M}_j)},
\end{equation}
where $\mathcal{B}_{ij} = p(d|\mathcal{M}_i)/p(d|\mathcal{M}_j)$ is the Bayes factor quantifying the relative support of the data for model $\mathcal{M}_i$ over $\mathcal{M}_j$, and $p(\mathcal{M}_i)/p(\mathcal{M}_j)$ is the prior odds. Therefore, the logarithmic Bayes factor $\ln \mathcal{B}_{ij}$ is computed as the difference between the logarithmic evidences of the models
\begin{equation}
\ln \mathcal{B}_{ij} \equiv \ln p(d|\mathcal{M}_i) - \ln p(d|\mathcal{M}_j).
\label{bayes_factor}
\end{equation}

The Jeffreys scale \cite{Kass:1995loi,Trotta:2008qt} establishes conventional thresholds for interpreting the strength of evidence based on logarithmic Bayes factors:
\begin{itemize}
\item $ |\ln \mathcal{B}_{ij}| < 1 $: Inconclusive evidence
\item $ 1 \leq |\ln \mathcal{B}_{ij}| < 2.5 $: Weak evidence
\item $ 2.5 \leq |\ln \mathcal{B}_{ij}| < 5 $: Moderate evidence
\item $ 5 \leq |\ln \mathcal{B}_{ij}| < 10 $: Strong evidence
\item $ |\ln \mathcal{B}_{ij}| \geq 10 $: Decisive evidence
\end{itemize}

Note that $\ln \mathcal{B}_{ij} > 0$ indicates a preference for model $\mathcal{M}_i$ over $\mathcal{M}_j$, while $\ln \mathcal{B}_{ij} < 0$ indicates the opposite. Here, we use publicly available code
{\tt MCEvidence}\footnote{\url{https://github.com/yabebalFantaye/MCEvidence}.} \cite{Heavens:2017hkr,Heavens:2017afc} to compute the Bayes factor of the models.

\begin{figure*}[htbp]
\begin{center}
\includegraphics[width=18cm]{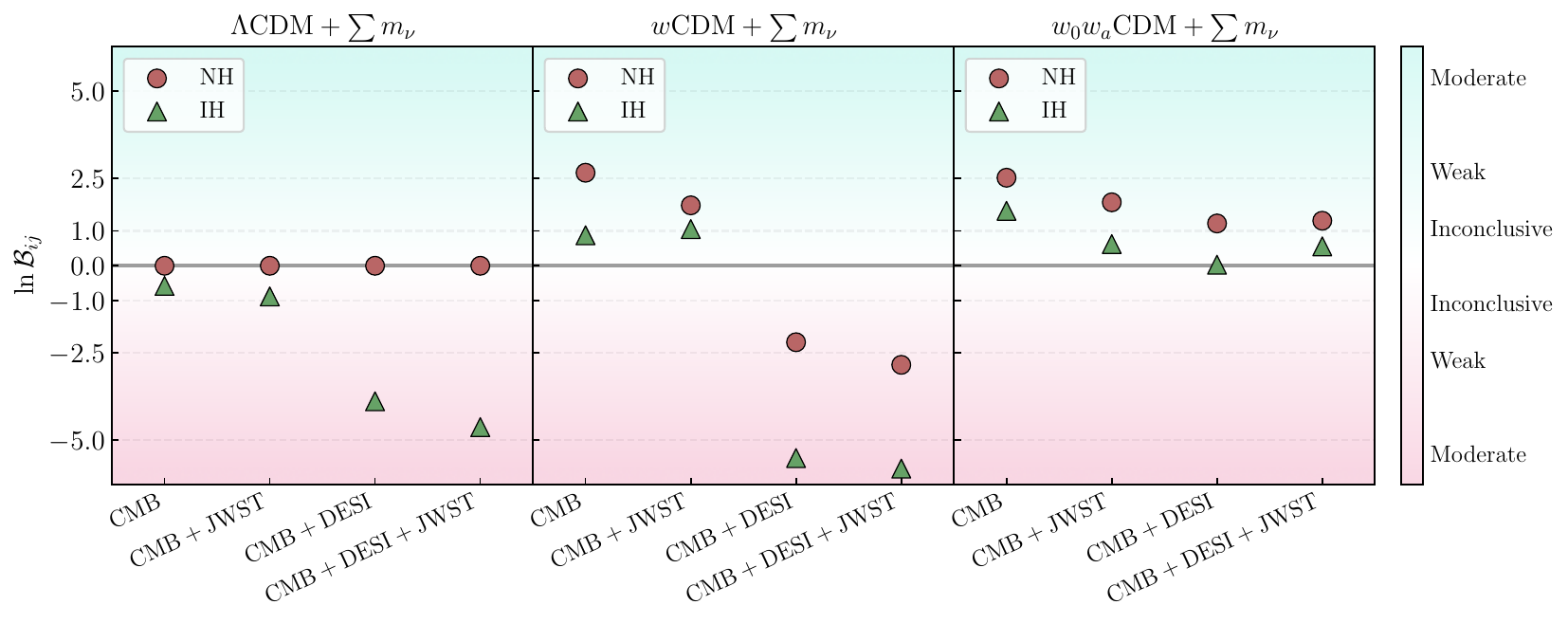}
\end{center}
\caption{The Bayes factors, $\ln \mathcal{B}_{ij}$, for three cosmological models, $\Lambda{\rm CDM}+\sum m_{\nu}$, $w{\rm CDM}+\sum m_{\nu}$, and $w_0w_a{\rm CDM}+\sum m_{\nu}$, derived from JWST, DESI, and CMB data. Filled circles and triangles represent the NH and IH cases, respectively. The background gradient illustrates the Jeffreys scale, with progressively deeper tones corresponding to stronger evidence. In summary, for each data combination, more positive values of $\ln \mathcal{B}_{ij}$ indicate that the model is more strongly favored by the data, whereas more negative values indicate that it is disfavored.
}
\label{figure3}
\end{figure*}

\section{Results and Discussions}\label{sec3}

In this section, we report neutrino mass measures derived from JWST, DESI, and CMB data. We show the cosmological constraints of $\Lambda$CDM+$\sum m_{\nu}$, $w$CDM+$\sum m_{\nu}$, and $w_0w_a$CDM+$\sum m_{\nu}$ models in Fig.~\ref{figure1} and Fig~\ref{figure2}. Table~\ref{tab3} summarizes the constraint results of cosmological parameters. And the comparison of different models based on Bayes factors is shown in Fig.~\ref{figure3}.

In Fig.~\ref{figure1}, we shows the one-dimensional marginalized posterior distributions of $\sum m_{\nu}$ for the $\Lambda\mathrm{CDM}+\sum m_{\nu}$, $w\mathrm{CDM}+\sum m_{\nu}$, and $w_{0}w_{a}\mathrm{CDM}+\sum m_{\nu}$ models, based on the CMB, CMB+JWST, CMB+DESI, and CMB+DESI+JWST combinations. First, focus on the NH case, the NH case is chosen as the benchmark because it represents the minimal $\sum m_{\nu}$ allowed by oscillation experiments, making it particularly sensitive to cosmological measures. Across all three models, JWST consistently tightens the upper limit of $\sum m_{\nu}$, with the most noticeable relative improvement obtained by CMB+JWST compared to CMB alone. For example, in $\Lambda$CDM, the upper limit of $\sum m_{\nu}$ in NH case improves from $0.243~\mathrm{eV}$ to $ 0.218~\mathrm{eV}$ ($10.3\%$), while in $w$CDM and $w_0w_a$CDM the corresponding improvements are from $0.286~\mathrm{eV}$ to $0.255~\mathrm{eV}$ ($10.8\%$) and from $0.296~\mathrm{eV}$ to $0.273~\mathrm{eV}$ ($7.8\%$), respectively.
When using CMB+DESI+JWST data, JWST continues to yield modest improvements for NH case, tightening the upper limits of $\sum m_{\nu}$ in $\Lambda$CDM, $w$CDM, and $w_0w_a$CDM by $0.9\%$, $3.2\%$, and $10.2\%$ compared with CMB+DESI. The slight improvement in $\Lambda$CDM is primarily due to the fact that this model is already close to the lower limit of $\sum m_{\nu}$. Notably, in the most flexible $w_0w_a$CDM model, which allows the largest parameter space among the three scenarios considered, the CMB+DESI+JWST combination still delivers a relatively tight limit of $\sum m_{\nu} < 0.167~\mathrm{eV}$ ($2\sigma$).

Similarly, in Fig.~\ref{figure1}, we continue to compare different neutrino mass hierarchies (DH, NH, and IH) with the $w_0w_a$CDM and using the CMB, CMB+JWST, CMB+DESI, and CMB+DESI+JWST data. For CMB alone, incorporating JWST yields appropriate improvements across all hierarchies: for DH case, the upper limit of $\sum m_{\nu}$ decreases from $0.247~\mathrm{eV}$ to $0.221~\mathrm{eV}$ ($10.5\%$), for the NH case, upper limit of $\sum m_{\nu}$ decreases from $0.296~\mathrm{eV}$ to $0.273~\mathrm{eV}$ ($7.8\%$), and for the IH case, upper limit of $\sum m_{\nu}$ decreases from $0.309~\mathrm{eV}$ to $0.277~\mathrm{eV}$ ($10.4\%$). When added to the CMB+DESI combination, JWST data still yields certain gains: for DH case, the upper limit of $\sum m_{\nu}$ tightens from $0.154~\mathrm{eV}$ to $0.145~\mathrm{eV}$ ($5.8\%$), for NH case, the upper limit of $\sum m_{\nu}$ tightens from $0.186~\mathrm{eV}$ to $0.167~\mathrm{eV}$ ($10.2\%$), and for IH case, the upper limit of $\sum m_{\nu}$ tightens from $0.195~\mathrm{eV}$ to $0.184~\mathrm{eV}$ ($5.6\%$). These results underscore that JWST high-redshift galaxy measurements are complementary to both CMB and low-redshift BAO data.

\begin{table}[t]
\renewcommand\arraystretch{1.7}
\centering
\caption{The summary of $\ln \mathcal{B}_{ij}$ values using current observational datasets. Note that $\ln \mathcal{B}_{ij}$ is computed with $p(d|\mathcal{M}_j)$ in Eq.~(\ref{bayes_factor}), taken to be the Bayesian evidence of the $\Lambda\mathrm{CDM}+\sum m_{\nu}({\rm NH})$ model.}
\label{tab4}
\resizebox{0.48\textwidth}{!}{%
\begin{tabular}{c@{\hspace{6mm}}cccccc}
\hline\hline
& \multicolumn{2}{@{\hspace{-10mm}}c}{\qquad$\Lambda\mathrm{CDM}+\sum m_{\nu}$\qquad} 
& \multicolumn{2}{@{\hspace{-5mm}}c}{~~~~~$w\mathrm{CDM}+\sum m_{\nu}$~~~~~} 
& \multicolumn{2}{@{\hspace{-8mm}}c}{~~~~$w_0w_a\mathrm{CDM}+\sum m_{\nu}$~~~~} \\
\cmidrule[0.5pt](l{-10pt}r{2pt}){2-3} 
\cmidrule[0.5pt](l{2pt}r{15pt}){4-5} 
\cmidrule[0.5pt](l{-10pt}r{10pt}){6-7}
&  NH & IH &  NH & IH &  NH & IH \\
\hline
\multicolumn{7}{l}{\small\textbf{CMB}} \\
$\ln\mathcal{B}_{ij}$ &   $0$ & $-0.58$ &  $2.67$ & $0.88$ &  $2.52$ & $1.58$ \\
\hline
\multicolumn{7}{l}{\small\textbf{CMB+JWST}} \\
$\ln\mathcal{B}_{ij}$ &   $0$ & $-0.88$ &  $1.73$ & $1.05$ &  $1.82$ & $0.62$ \\
\hline
\multicolumn{7}{l}{\small\textbf{CMB+DESI}} \\
$\ln\mathcal{B}_{ij}$ &   $0$ & $-3.88$ &  $-2.19$ & $-5.51$ &  $1.21$ & $0.03$ \\
\hline
\multicolumn{7}{l}{\small\textbf{CMB+DESI+JWST}} \\
$\ln\mathcal{B}_{ij}$ &   $0$ & $-4.62$ &  $-2.84$ & $-5.81$ &  $1.29$ & $0.56$ \\
\hline
\end{tabular}
}
\end{table}

The SFE parameter determines the abundance of high-redshift galaxies, which is sensitive to the suppression of small-scale structures caused by massive neutrinos. As shown in Fig.~\ref{figure2}, variations in cosmological parameters across the range allowed by CMB+DESI+JWST do lead to changes in the SFE parameter; however, these variations are not substantial enough to qualitatively alter the main conclusions. The data yield a $2\sigma$ lower limit of $f_{*,10} \gtrsim 0.146-0.161$. This means that very low SFE parameter ($f_{*,10} \lesssim 0.1$) values are disfavored, although a wide range of higher efficiencies remains allowed. Furthermore, SFE is sensitive to assumptions about stellar-mass estimation, the initial mass function, sample completeness and contamination, and the parametric form of the halo-stellar relation. Meanwhile, current JWST high-redshift samples are still limited by statistics and potential systematics in photometric redshifts and mass-to-light ratios. Therefore, while the JWST data already set interesting limits on SFE, a more robust measurement will require spectroscopic confirmation and improved control of these systematics \cite{Labbe:2022ahb,Liu:2023qkf}.

Finally, we employ Bayesian evidence to compare different models as shown in Fig.~\ref{figure3} and summarized the values of $\ln \mathcal{B}_{ij}$ in Table~\ref{tab4}. For each data combination, larger values of $\ln \mathcal{B}_{ij}$ indicate a stronger preference for model $\mathcal{M}_i$ over model $\mathcal{M}_j$. Crucially, we restrict our interpretation to the physically motivated NH and IH cases, excluding the DH case from comparative analysis due to its theoretical inconsistency with neutrino oscillation parameters. For $w_0w_a$CDM+$\sum m_{\nu}$(NH) model, the values of $\ln \mathcal{B}_{ij}$ are 2.52, 1.82, 1.21, and 1.29 using the CMB, CMB+JWST, CMB+DESI, and CMB+DESI+JWST data, respectively. For all data, the $w_0w_a$CDM model is consistently weakly favored by Bayesian evidence, with the NH case also weakly favored.

\section{Conclusion}\label{sec4}

In this work, we perform a joint analysis of JWST, DESI, and CMB data to measure $\sum m_{\nu}$ within three dark energy models, while accounting for three neutrino mass hierarchies. We further report the Bayesian evidence to assess the statistical preference among competing cosmological scenarios.

In the $w_0w_a$CDM model with NH case, compared with CMB alone, the inclusion of JWST data tightens the $2\sigma$ upper limit of $\sum m_{\nu}$ from $0.296~\mathrm{eV}$ to $0.273~\mathrm{eV}$ (a $7.8\%$ improvement). Using the joint CMB+DESI+JWST dataset, we obtain $\sum m_{\nu} < 0.167~\mathrm{eV}$ ($2\sigma$), representing a 10.2\% improvement over the CMB+DESI limit. These results demonstrate that JWST observations of high-redshift massive galaxy candidates provide additional constraining power on the absolute neutrino mass, particularly in dynamical dark energy models with complex parameter degeneracies. In addition, JWST data yield a $2\sigma$ lower limit on the SFE parameter, $f_{*,10} \gtrsim 0.146-0.161$. However, this constraint remains subject to significant model dependence and systematic uncertainties. Finally, Bayesian evidence further shows that the CMB+DESI+JWST data marginally weakly favors the $w_0w_a$CDM+$\sum m_{\nu}$(NH) framework, with a Bayes factor of 1.29 relative to $\Lambda$CDM+$\sum m_{\nu}$(NH). 

In summary, jointly analyzing JWST high-redshift and DESI low-redshift data provides useful constraints on neutrino mass. In the coming years, forthcoming larger and more precise JWST datasets combined with the complete DESI dataset, as well as future CMB observations from the CMB-S4 \cite{CMB-S4:2016ple}, could be utilized to measure $\sum m_{\nu}$, potentially uncovering more unexpected and intriguing results.

\section*{ACKNOWLEDGMENTS}
We thank Jian-Qi Liu, Jia-Le Ling and Yi-Min Zhang for their helpful discussions. This work was supported by the National SKA Program of China (Grants No. 2022SKA0110200 and No. 2022SKA0110203), the National Natural Science Foundation of China (Grants No. 12533001, No. 12575049, and No. 12473001), the China Manned Space Program (Grant No. CMS-CSST-2025-A02), and the National 111 Project (Grant No. B16009). This work uses publicly available JWST data from the CEERS program obtained via MAST.

\section*{DATA AVAILABILITY}
The data that support the findings of this article are openly available~\cite{Zhou_ScienceDB_28617}.

\bibliography{main} 

\end{document}